\documentclass[pra,showpacs,showkeys,preprintnumbers,amsmath,amssymb,superscriptaddress,twocolumn]{revtex4-1}
\usepackage[english]{babel}
\usepackage{makeidx} 
\usepackage{graphicx} 
\usepackage{dcolumn} 
\usepackage{array} 
\usepackage{amssymb} 
\usepackage{amsmath}
\usepackage{textcomp}
\usepackage{multirow}
\usepackage{subfigure}
\usepackage{eucal}
\usepackage{mathrsfs}
\usepackage[all]{xy}
\usepackage{epstopdf}

\usepackage{color}

\usepackage{float} 
\usepackage{amsmath} 
\usepackage{amsfonts} 
\usepackage{bm} 

\begin{document}

\title{Quantum correlations and Bell's inequality violation in a Heisenberg spin dimer via neutron scattering} 

\author{C. Cruz\email{clebsonscruz@yahoo.com.br}} \affiliation{Instituto de F\'{i}sica, Universidade Federal Fluminense, Av. Gal. Milton Tavares de Souza s/n, 24210-346 Niter\'{o}i, Rio de Janeiro, Brazil.}

\date{\today}

\begin{abstract}
The characterization of quantum information quantifiers has attracted a considerable attention of the scientific community, since they are a useful tool to verify the presence of quantum correlations in a quantum system. In this context, in the present work we show a theoretical study of some quantifiers, such as entanglement witness, entanglement of formation, Bell's inequality violation and geometric quantum discord as a function of {the diffractive} properties of neutron scattering. We provide one path toward {identifying} the presence of quantum correlations {and quantum nonlocality} in a molecular magnet as a Heisenberg spin-$1/2$ dimer, by diffractive properties typically obtained via neutron scattering experiments.
\end{abstract}
\maketitle

\section{Introduction}

The study of quantum correlations has been {subject} of numerous investigations in the last {few} years, since it is a remarkable resource in quantum information science. In this {regard}, quantum information quantifiers \cite{nielsen,horodecki,horodecki1996separability,wiesniak2005magnetic,souza2009entanglement,wootters,hill,brukner2006crucial,landau1987violation,souza2008nmr,zurek,vedral4,liu,ma,adesso2,adesso,cruz,sarandy,paula,sarandy3,luo,datta,vedral,paula,sarandy3,huang2013quantum,huang2014scaling} are {useful tools} to verify the presence of quantum correlations in a quantum system. In spite of that, the detection of quantum correlations is a {difficult task,} theoretically and experimentally speaking \cite{cruz,liu,adesso,adesso2,girolami,girolami2,luo,huang2014computing}.
Nowadays, it {is} understood that quantum correlations can be quantified through the {measurements} of some macroscopic properties of magnetic systems \cite{cruz,yuri,yuri2,liu}. 

The study of the magnetic properties of molecular materials{is} typically done approximating magnetic parameters of a Hamiltonian model by the fit of some thermodynamic properties, such as magnetic susceptibility, internal energy and specific heat  \cite{mario,cruz,haraldsen2005neutron,leite2015heptacopper,esteves2014new,leite2015heptacopper}. In this context, {correlation functions have great importance in describing these properties; in addition, they} can be directly measurable, e.g., in neutron scattering experiments via \textit{structure factors}. Structure factors can be defined {as two-point correlations \cite{krammer} and are} widely used to describe the crystal structure of molecular systems ruled by Hamiltonians \cite{krammer,tennant2003neutron,huberman2005two}, e.g., Heisenberg models \cite{mario}. 

Quantum information quantifiers are expressed in terms of statistical correlation functions \cite{yuri,yuri2,cruz}, due to the fact that these functions are {present in} the elements of the density matrix of the quantum system, linking their macroscopic properties with the quantum ones. Therefore, it is possible {to} quantify the presence of quantum correlations in a system via structure factors \cite{cramer,krammer,liu,marty}, since these factors are directly associated to the correlation functions; {thus} allowing the measurement of quantum information quantifiers by neutron scattering experiments.

In this scenario, the present work shows analytical expressions for the entanglement witness, entanglement of formation, Bell's inequality violation{,} and geometric quantum discord, based on {the} Schatten 1-norm as a function of quantities typically obtained in neutron scattering via {a} scalar structure factor. Our results provide one path toward {identifying} the presence of quantum correlations {and quantum nonlocality} in a molecular magnet such as a Heisenberg spin-$1/2$ dimer, by diffractive properties. This is an alternative way to describe the quantum properties of a sample material via neutron scattering experiments, without {making} any assumption about their macroscopic quantities, leading to promising applications in quantum information science.

\section{Neutron scattering for a Heisenberg spin dimer}

The study of molecular magnetic materials is typically done through the approach of {the} magnetic parameters of a {Hamiltonian model}, by the fit of some thermodynamic properties, e.g., magnetic susceptibility, internal energy{,} and specific heat  \cite{mario,cruz,haraldsen2005neutron,leite2015heptacopper,esteves2014new,leite2015heptacopper}. For a given Hamiltonian model, one can evaluate the inelastic structure factor, which allows a sensitive test of the assumed model, since their properties are affected by the relative positions 
of the metallic centers of a sample material  \cite{haraldsen2005neutron}. 

Let us consider a molecular magnet as {an} interacting pair of spin-$1/2$ ruled by the Heisenberg-{Dirac}-Van Vleck Hamiltonian,
\begin{eqnarray}
\mathcal{H}&=&-J \vec{S}_1\cdot \vec{S}_2 ~.
\label{hei}
\end{eqnarray}
This is an ideal realization of a two qubit system; and therefore, a promising platform in the quantum information processing. 

Once Eq. (\ref{hei}) is invariant under spin rotation, the total spin $s=s_1+s_2$ is a good quantum number \cite{haraldsen2005neutron,mario}. From the Clebsch-Gordon series the spectrum consists in an $s=1$ triplet and $s=0$ singlet \cite{mario}. Diagonalizing it, we obtain the energy eigenvalues $E_s$ and eigenvectors $\vert s, m_s \rangle$ \cite{haraldsen2005neutron,mario}:
\begin{align}
&E_{s=1}=\frac{1}{4}J  \\
&E_{s=0}=-\frac{3}{4}J \\
&\vert s=1,m_s=+1\rangle =\vert 00\rangle \\
&\vert s=1,m_s=0\rangle =\frac{1}{\sqrt{2}}\left( \vert 01\rangle + \vert 10\rangle\right) \\
&\vert s=1,m_s=-1\rangle =\vert 11\rangle \\
&\vert s=0, m_s=0\rangle =\frac{1}{\sqrt{2}}\left( \vert 01\rangle {-} \vert 10\rangle\right) ~.
\end{align}

In magnetic neutron scattering, neutrons interact magnetically with the atoms of the target sample. From the Van Hove formalism \cite{hove}{, the} partial differential cross section of {an} incident neutron in a magnetic system with initial state $\vert\psi_{i}\rangle$ is expressed in terms of time-dependent correlation functions
{
\begin{eqnarray}
\frac{d^2\sigma}{d\omega d\Omega} &=&\frac{1}{Z} \left( \frac{g_{i, \alpha}\gamma r_{0}}{2}\right)^2  \nonumber\\ \nonumber
&& \times \sum_{\alpha ,\beta , l, m} \left( \delta_{\alpha , \beta} -\frac{\vec{q_\alpha}\vec{q_\beta}}{\vert \vec{q}\vert^2}\right) \left[ F_l(\vec{q})g_{l, \alpha}\right]^{*}\left[  F_m(\vec{q})g_{m, \beta}\right] \nonumber \\ 
&&\times\int \frac{dt}{2\pi}e^{i\left[ \omega t + \vec{q}\cdot \left( \vec{r_l} - \vec{r_m}\right)\right]}\langle\psi_{i}\vert S_l^\alpha S_m^\beta(t)\vert\psi_{i}\rangle,
\label{eq:18}
\end{eqnarray}
where the sum run over all magnetic sites $l$ and $m$, which are the l-th and m-th spins with position vectors $\vec{r_l}$ and $\vec{r_m}$, respectively. Furthermore, $\gamma$ is the neutron magnetic moment; $r_0=e^2 /m_ec^2$ is the classical electron radius;  $F_i(\vec{q})$ is the magnetic form factor; $\alpha$,$\beta=x,y,z$;  $\omega$ is the energy transferred to the target magnetic system; $\vec{q}=\vec{k}_{out}-\vec{k}_{in}$ is the difference between the final and the initial wave vectors (scattering vector); and finally, $S_l^\alpha$ are the spin operators.
}
{Thus, the differential cross section is proportional to  the neutron scattering structure factor tensor that is written in terms of the pairwise correlation function\cite{marty,zali,liu,cramer}:
}
\begin{eqnarray}
\mathcal{S}^{(\alpha , \beta)}(\vec{q},\omega)=\int \frac{dt}{2\pi}  \sum_{l, m} e^{i\left[ \omega t + \vec{q}\cdot \left( \vec{r_l} - \vec{r_m}\right)\right]}\langle\psi_{i}\vert S_l^\alpha S_m^\beta(t)\vert\psi_{i}\rangle. \nonumber \\
\label{integral}
\end{eqnarray}

Due to the transitions between the discrete energy levels, Eq. (2)-(3), the time integral shown in Eq. (\ref{integral}) gives a delta function $\delta\left( E_{f} - E_{i} - \hbar\omega\right)$, where $\hbar\omega$ is the transfer energy  \cite{haraldsen2005neutron}. 
{Integrating Eq.(\ref{integral}) over energies, we obtain the integrated structure factor \cite{krammer,marty,cramer}
{
\begin{equation}
\bar{\mathcal{S}}(\vec{q})=\sum_{\alpha,\beta}{\mathcal{S}}^{\alpha\beta}(\vec{q}) = \sum_{\alpha,\beta}\sum_{l,m} e^{i\vec{q}\cdot \left( \vec{r_l} - \vec{r_m}\right)} \langle S_l^\alpha S_m^\beta\rangle.
\label{static}
\end{equation}
}

{Therefore, for the specific case of a Heisenberg spin dimer, Eq.(\ref{hei}), one can define the so-called exclusive structure factor as a function of the scattering vector $\vec{q}$ \cite{haraldsen2005neutron}, for the excitation of the final states in the magnetic multiplet $\vert\psi_{f}\rangle$, Eq.(4)-(7), from the given initial state $\vert\psi_{i}\rangle$ 
\begin{eqnarray}
\mathcal{S}^{(\alpha , \beta)}_{fi}(\vec{q})=\sum_f \langle \psi_{i} \vert \mathcal{U}_{\alpha}^\dagger(\vec{q}) \vert \psi_{f}\rangle \langle \psi_{f}\vert \mathcal{U}_{\beta}(\vec{q})\vert \psi_{i}\rangle,
\label{exclusive}
\end{eqnarray}
where 
\begin{eqnarray}
\mathcal{U}_{\alpha,\beta}(\vec{q})= \sum_{\vec{r}_l} S_l^{(\alpha,\beta)} e^{i\vec{q}\cdot r_l},\end{eqnarray}
and the sum taken over all magnetic ions in a unit cell \cite{haraldsen2005neutron,liu,marty}.
}

{
For the $\vert s=0, m_s=0\rangle$  initial state (antiparallel magnetic alignment), Eq.(7), only $s=1$ final states are excited, Eq.(4)-(6). Thus, one can define the scalar neutron scattering structure factor, $\mathcal{S}(\vec{q})$, for the Heisenberg $1/2$-spin dimer \cite{haraldsen2005neutron} by
\begin{eqnarray}
\mathcal{S}^{(\alpha , \beta)}_{fi}(\vec{q})= \delta_{\alpha,\beta}\mathcal{S}(\vec{q}).
\label{scalar}
\end{eqnarray}
}
{Using the eigenvectors given by Eq.(4)-(7) in Eq.(\ref{exclusive})}, we evaluate the inelastic neutron scattering intensities, that {are} given by structure factors \cite{haraldsen2005neutron} for the Heisenberg spin-$1/2$ dimer, Eq. (\ref{hei}). Thus, we obtain the scalar neutron scattering structure factor as calculated in reference \onlinecite{haraldsen2005neutron}:
\begin{equation}
\mathcal{S}(\vec{q})= \frac{1}{2}\left[1 - cos(\vec{q}\cdot \left( \vec{r_1} - \vec{r_2}\right))\right].
\label{structure}
\end{equation}

Thus, given the Hamiltonian model, it is possible to predict their scalar structure factor, which can be compared to the neutron scattering experiments results \cite{haraldsen2005neutron}.

\section{Quantum information-theoretic quantifiers in a Heisenberg spin dimer}

In this section, we will present the analytical expressions for quantum information quantifiers, such as entanglement witness, entanglement of formation, Bell's inequality violation and geometric quantum discord, based on {the} Schatten 1-norm as a function of quantities typically obtained in neutron scattering via scalar structure factor, allowing their measurement via neutron scattering experiments.

\subsection{Spin-Spin Correlation Function}

{Correlation functions have} a great importance in statistical physics and quantum mechanics, allowing {us} to find  different properties of a physical system. In addition, it can be directly measurable, e.g., in scattering experiments \cite{yuri2}. Quantum information quantifiers are expressed in terms of statistical correlation functions \cite{yuri,yuri2,cruz}, due to the fact that these functions are {present in} the elements of the density matrix of the quantum system, linking their macroscopic {and structural} properties with the quantum ones.

{
For the system ruled by the Hamiltonian, Eq.(\ref{hei}), from the Eq.(\ref{static}) and Eq.(\ref{scalar}) the spin-spin correlation function can be extracted and written in terms of integrated structure factor
\begin{eqnarray}
\mathcal{C} &=& \langle S_1^\alpha S_2^\alpha \rangle = e^{-i\vec{q}\cdot \left( \vec{r_1} - \vec{r_2}\right)} \mathcal{S}^{\alpha\alpha}(\vec{q}) ,\label{eq:23}
\end{eqnarray}
where $\alpha=x,y,z$.
}

{
Thus, spin-spin correlation function can be accessed through diffractive properties obtained via neutron scattering experiments, without making any assumption about the macroscopic properties of the measured system or the external conditions under which the neutrons are scattered.
}

{Using Eq.(\ref{structure}) spin-spin correlation function can be written as a function of the the scattering vector, $\vec{q}=\vec{k}_{out}-\vec{k}_{in}$, i.e., the difference between the final and the initial wave vectors, and the the difference between the position vectors $\vec{r_1}$ and $\vec{r_2}$ of the metallic centers.}
\begin{eqnarray}
\mathcal{C}&=& \frac{e^{-i\vec{q}\left( \vec{r_1}-\vec{r_2}\right)}}{2}\left[1 - cos(\vec{q}\cdot \left( \vec{r_1} - \vec{r_2}\right))\right] ~,
\label{spin}
\end{eqnarray}
this last ranges from $-1\leq \mathcal{C} \leq 0$ for antiparallel magnetic alignment ($J<0$ - entangled ground state) and  $0< \mathcal{C} \leq 1/3$ for parallel magnetic alignment ($J>0$ - separable ground state) \cite{cruz,yuri,yuri2}.
\begin{figure}[ht]
\centering
\includegraphics[scale=0.33]{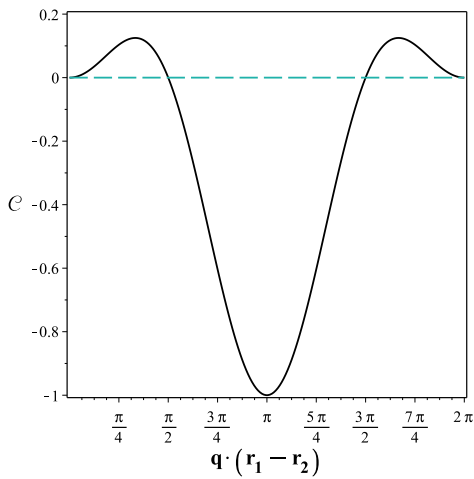}
\caption{Spin-spin correlation as a function of $\vec{q}\cdot\left(\vec{r_1} - \vec{r_2}\right)$, Eq.(\ref{spin}). The dashed (green) horizontal line separates the regions with antiparallel magnetic alignment ($\pi/2 \leq  \vec{q} \cdot \left( r_1 - r_2 \right) \leq  3\pi /2$) and parallel magnetic alignment ($0 < \vec{q} \cdot \left( r_1 - r_2 \right) < \pi/2 $ and $3\pi /2 < \vec{q} \cdot \left( r_1 - r_2 \right) < 2\pi$ ). }
\label{fig:spin}
\end{figure}

The spin-spin correlation function, Eq. (\ref{spin}), is depicted in Fig. (\ref{fig:spin}) as a function of the scattering vector $\vec{q}$ times the distance between the metallic centers of a sample material $\vec{r_1} - \vec{r_2}$, {which} can be obtained by neutron scattering experiments. It is worth noting that, the location of the zero point correlation can provide a convenient estimate for which distances {in} the system can be found in an antiparallel magnetic alignment ($\pi q^{-1}/2 \leq   r_1 - r_2 \leq  3\pi q^{-1}/2$) or a parallel magnetic alignment ($0 < r_1- r_2 < \pi q^{-1}/2 $ and $3\pi q^{-1}/2 < r_1 - r_2 < 2\pi q^{-1} $ ), without any assumption about macroscopic quantities, such as the temperature, magnetic field, magnetic susceptibility, internal energy or specific heat. 
This result has a great importance to guide us through the study of the quantum information quantifiers in a Heisenberg spin dimer, since these quantifiers depend directly {on} the spin-spin correlation function.

\subsection{Entanglement Witness}

The detection of entanglement is usually done using an observable which identifies the presence of entanglement in a quantum system \cite{horodecki,horodecki1996separability,wiesniak2005magnetic,souza2009entanglement}. This observable, {the so-called} \textit{entanglement witness} ($\mathcal{W}$),  has a negative expectation value whether the system is in an entangled quantum state ($\mathcal{W}<0$) and  {otherwise} positive.
However, the positive expectation value  does not {imply} the presence of separable quantum states.
{Recently magnetic susceptibility was proposed as a thermodynamical entanglement witness \cite{horodecki,horodecki1996separability,wiesniak2005magnetic,souza2009entanglement}. For a system in which $[\mathcal{H},S_z]$ the average magnetic susceptibility in a complete separable state satisfies \cite{wiesniak2005magnetic,souza2009entanglement}
\begin{eqnarray}
\bar{\chi}=\frac{\chi_x +\chi_y + \chi_z}{3}\leqslant \frac{(g\mu_B)^2NS}{3k_BT},
\end{eqnarray}
where $N$ is the number of magnetic ions with spins-$S$, $k_B$ is the Boltzmann constant, $\mu_B$ is the Bohr magneton, $g$ is the Land\'{e} factor and $\bar{\chi}$ is the average of the magnetic susceptibility.}

As calculated {by Wie\'{s}niak, Vedral and Brukner in the} reference \onlinecite{wiesniak2005magnetic}, the entanglement witness can be calculated in terms of {the average magnetic susceptibility:}
\begin{equation}
\mathbb{W} = \frac{3k_BT\bar{\chi}}{(g\mu_B)^2NS} -1.
\label{witness}
\end{equation}

In reference \onlinecite{brukner2006crucial}, the magnetic susceptibility of {an} antiferromagnetic spin-1/2 chain is compared to the correlation function measured by neutron diffraction \cite{soares2009entanglement}. {The pairwise average magnetic susceptibility  must be written as a function of the pairwise correlation, Eq.(\ref{spin}), as
\begin{equation}
\bar{\chi} (T)=\frac{N(g\mu_B)^2(1+\mathcal{C})}{2 k_B T }.
\label{spinsus}
\end{equation}
}

Thus, we can {analytically calculate the entanglement witness for a molecular magnet, Eq.(\ref{witness})}, such as a Heisenberg spin-$1/2$ dimer, in terms of their scalar structure factor using Eq.(\ref{spin}), (\ref{witness}) and (\ref{spinsus})
\begin{eqnarray}
\mathbb{W} &=& 2 + 3 e^{-i\vec{q}\left( \vec{r_1}-\vec{r_2}\right)}\mathcal{S}(\vec{q}) \nonumber \\
 &=& 2 + \frac{3 e^{-i\vec{q}\left( \vec{r_1}-\vec{r_2}\right)}}{2}\left[ 1 - cos(\vec{q}\cdot \left( \vec{r_1} - \vec{r_2}\right))\right].
\label{witness2}
\end{eqnarray}
\begin{figure}[ht]
\centering
\includegraphics[scale=0.325]{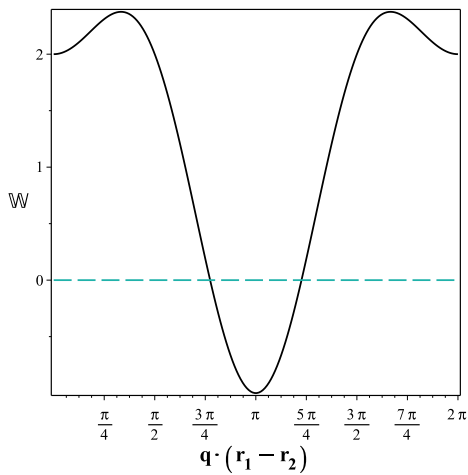}
\caption{Entanglement witness of a Heisenberg spin dimer, Eq.(\ref{witness2}). The dashed (green) horizontal line separates the region where the witness has a negative expectation value, i.e., the system is in an entangled quantum state $2.45 < \vec{q} \cdot \left( \vec{r_1} - \vec{r_2} \right) < 3.85$. }
\label{fig:witness}
\end{figure}

In Fig. \ref{fig:witness}, we show the entanglement witness for a Heisenberg spin dimer, Eq. (\ref{witness2}), as a function of $\vec{q} \cdot \left( r_1 - r_2 \right)$. The witness has a negative expectation at {the range} $2.45 < \vec{q} \cdot \left( \vec{r_1} - \vec{r_2} \right) < 3.85$, revealing the presence of entangled states. This result is compatible  to the last ones, since in this band the system is found in an antiparallel magnetic alignment (entangled ground state), as can be seen in Fig. \ref{fig:spin}. Thus, we provide one way to identify the presence of entanglement in a molecular magnet, such as a Heisenberg spin dimer, by diffractive properties obtained via neutron scattering experiments, without {making} any assumption about their macroscopic quantities. 

\subsection{Entanglement of Formation}

In order to quantify the amount of entanglement in the Heisenberg spin-$1/2$ dimer and make a comparison with the entanglement witness, we will adopt the measurement of \textit{entanglement of formation} defined by  \cite{wootters,hill}:
\begin{eqnarray}
\mathbb{E}=-\Gamma_{+}\log_2 \left(\Gamma_{+}\right)-\Gamma_{-}\log_2 \left(\Gamma_{-}\right),
\label{eq:9}
\end{eqnarray}
with
\begin{equation}
\Gamma_{\pm}=\frac{1\pm\sqrt{1-\mathbb{C}^2}}{2},
\end{equation}
where $\mathbb{C}$ is the concurrence \cite{wootters,hill,nielsen,horodecki}.  The concurrence can be  written as a function of the scalar structure factor, Eq.(\ref{structure}), in terms of spin-spin correlation function, Eq.(\ref{spin}), as follows:
\begin{equation}
\mathbb{C} = \mbox{max}\left[ 0, -\frac{1}{2}\left( 1 + 3 e^{-i\vec{q}\left( \vec{r_1}-\vec{r_2}\right)}\mathcal{S}(\vec{q})\right) \right].
\label{eq:10}
\end{equation}

The equation above shows that the concurrence of a Heisenberg spin-$1/2$ dimer is also related {to the} diffractive properties, which can be obtained by neutron scattering experiments.
\begin{figure}[ht]
\centering
\subfigure[]{\includegraphics[scale=0.33]{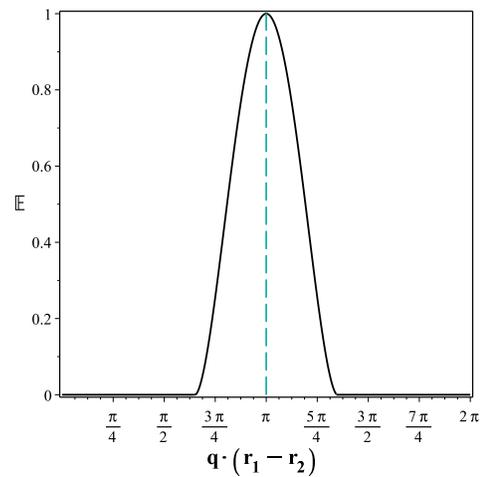}}\qquad
\subfigure[]{\includegraphics[scale=0.33]{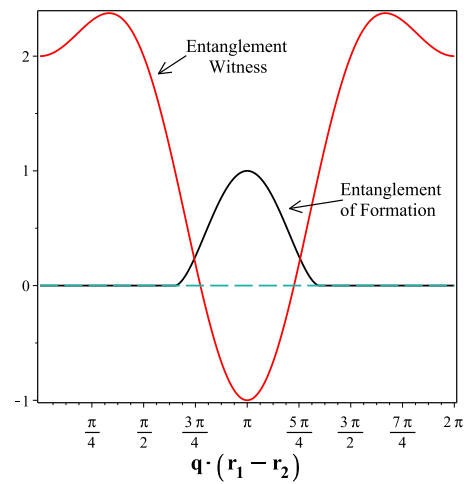}}
\caption{(a) Entanglement of formation as a function of $\vec{q}\cdot\left(\vec{r_1} - \vec{r_2}\right)$, Eq.(\ref{eq:9}). The dashed (green) vertical line highlights the maximum of entanglement, where the system is in an antiparallel magnetic alignment with an entangled pure state. (b) We make a comparison between the entanglement of formation and the entanglement witness. The dashed (green) horizontal line separates the region where the witness has a negative expectation value, i.e., the system is in an entangled quantum state.}
\label{fig:entanglement}
\end{figure}

In sequence, we shown in Fig. \ref{fig:entanglement} (a) the entanglement of formation as a function of the scattering vector $\vec{q}$ times the distance between the metallic centers $\vec{r_1} - \vec{r_2}$ of a Heisenberg spin dimer, Eq.(\ref{eq:9}). It is possible to identify a maximum of entanglement at $r_1 - r_2 = \pi q^{-1}$, where the system is found in an antiparallel magnetic alignment with an entangled pure state. It is worth {noting that} there {are} entangled states at $2.0 \lesssim \vec{q} \cdot \left( r_1 - r_2 \right) \lesssim 4.2$, above the band where the entanglement is identified on the entanglement witness. In Fig. \ref{fig:entanglement} (b), we make a comparison between the entanglement of formation and the entanglement witness. As can be seen, the positive expectation value of the witness{,} separated by the dashed (green) line, does not {imply} separability. However, the negative expectation value {necessarily implies in the} presence of entangled quantum states in the system. 

\subsection{Bell's Inequality Violation}

\textit{Bell's inequality violation} \cite{landau1987violation} has {an} important role in the quantum information theory, as a necessary and sufficient condition for the usefulness of quantum states in {the} communication complexity of protocols \cite{brukner2004bell}. For a Heisenberg spin-$1/2$ dimer, the Bell's inequality test \cite{souza2009entanglement,souza2008nmr} {is} related to the measurement of the Bell operator \cite{souza2009entanglement}
\begin{equation}
\mathbb{B}=\vec{n_1}\cdot\vec{S}\otimes\left( \vec{n_2}\cdot\vec{S} - \vec{n_4}\cdot\vec{S}\right) + \vec{n_3}\cdot\vec{S}\otimes\left( \vec{n_2}\cdot\vec{S} + \vec{n_4}\cdot\vec{S}\right),
\label{bell}
\end{equation}
where $\vec{n_1}\cdot\vec{S}$ is the projection of the spin on the direction $n$.

Using the set of directions $\lbrace\vec{n_1},\vec{n_2},\vec{n_3},\vec{n_4}\rbrace = \lbrace (0,0,1), (-1,0,-1)/\sqrt{2},(-1,0,0),(-1,0,1)/\sqrt{2} \rbrace$, the Eq. (\ref{bell}) becomes $\mathbb{B}=\sqrt{2}\left( S_x\otimes S_x + S_z\otimes S_z \right)$. Therefore 
\begin{eqnarray}
\vert\langle\mathbb{B}\rangle\vert &=& 2\sqrt{2}\vert \mathcal{S}(\vec{q})\vert \nonumber \\
&=& 2\sqrt{2}\left| 1 - cos(\vec{q}\cdot \left( \vec{r_1} - \vec{r_2}\right))\right|.
\label{bell2}
\end{eqnarray}

Thus, from Eq.(\ref{bell2}), it is possible to verify whether there is {a} Bell's inequality violation via neutron scattering experiments.
\begin{figure}[ht]
\centering
\includegraphics[scale=0.33]{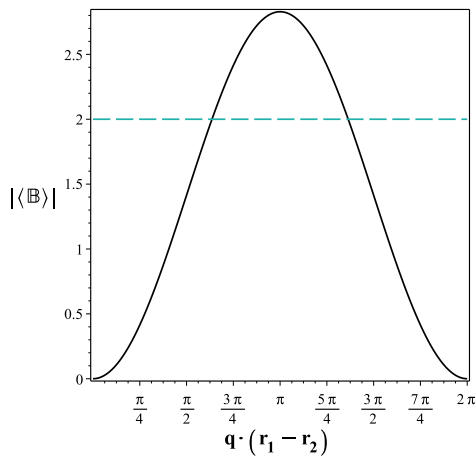}
\caption{Mean value of Bell operator as a function of $\vec{q}\cdot\left(\vec{r_1} - \vec{r_2}\right)$, Eq.(\ref{bell2}). The dashed (green) horizontal line separates the region where the Heisenberg dimer violates the Bell's inequality.}
\label{fig:bell}
\end{figure}

In Fig. \ref{fig:bell}, we shown the mean value of {the} Bell operator as a function of the scattering vector $\vec{q}$ times the distance between the metallic centers $\vec{r_1} - \vec{r_2}$ of a target sample of neutron scattering, Eq.(\ref{bell2}). As can be seen, the Bell's inequality is violated in $2.0 \lesssim \vec{q} \cdot \left( r_1 - r_2 \right) \lesssim 4.2$, where there is a maximum violation in $ r_1 - r_2 = \pi q^{-1} $, when the system is in an entangled pure state, see Fig. \ref{fig:entanglement} (a). It is compatible to the previous result obtained from the entanglement of formation, where we found entangled quantum states at the same band. 
Therefore, it is possible to verify {the} quantum nonlocality in a Heisenberg spin dimer by {diffractive} properties obtained in neutron scattering experiments.

\subsection{Geometric Quantum Discord}

Despite quantum entanglement {providing} one path toward {finding} pure quantum correlations, it does not encompass all quantum correlations in a system. The measurement of the total amount of quantum correlations has been called \textit{quantum discord} \cite{zurek,vedral4,liu,ma,adesso2,adesso,cruz,sarandy,paula,sarandy3,luo,datta,vedral}.

The calculation of quantum discord is a rather complicated task{,} even for  {a} two-qubit system, such as a Heisenberg spin-$1/2$ dimer \cite{cruz,paula}. This fact has stimulated alternative measurements of quantum information-theoretical quantifiers as the \textit{geometric quantum discord} \cite{cruz}. In this context, the geometric quantum discord, based on {the} Schatten 1-norm, is one path toward {achieving} a well-defined measurement of quantum correlations in a quantum system \cite{cruz} and it can be defined as
\begin{eqnarray}
\mathbb{Q}_{G}(\rho)=\min_{\omega}\Vert\rho - \rho_c\Vert,
\label{eq:7}
\end{eqnarray}
where $\Vert X\Vert=\mbox{Tr}\left[\sqrt{X^\dagger X}\right]$ is the 1-norm, $\rho$ is a given quantum state and $\omega$ is the set of closest classical-quantum states $\rho_c$ \cite{paula,sarandy3,cruz}, whose general form is given by:
\begin{equation}
\rho_c=\sum_{k}p_{k}\Pi_{k}^{\lbrace 1 \rbrace}\otimes\rho_{k}^{\lbrace 2 \rbrace},
\end{equation}
with $0 \leq {p_k} \leq 1$ and $\sum_{k} p_k = 1$; $\lbrace\Pi_{k}^{\lbrace 1 \rbrace}\rbrace$ denotes a set of
orthogonal projectors for subsystem {$1$}, and $\rho_{k}^{\lbrace 2 \rbrace}$ is a general reduced density operator for the subsystem $2$ \cite{paula,sarandy3}.

Thus, for a Heisenberg spin dimer, the geometric quantum discord, based on {the} Schatten 1-norm, is written as a function of the scalar structure factor as:
\begin{eqnarray}
\mathbb{Q}_{G}(\mathcal{S}) &=& \left| \frac{1}{4}\mathcal{S}(\vec{q}) \right| \nonumber \\
&=& \left| \frac{1 - cos(\vec{q}\cdot \left( \vec{r_1} - \vec{r_2}\right))}{4}\right| ~.
\label{geometric}
\end{eqnarray}
\begin{figure}[ht]
\centering
\subfigure[]{\includegraphics[scale=0.33]{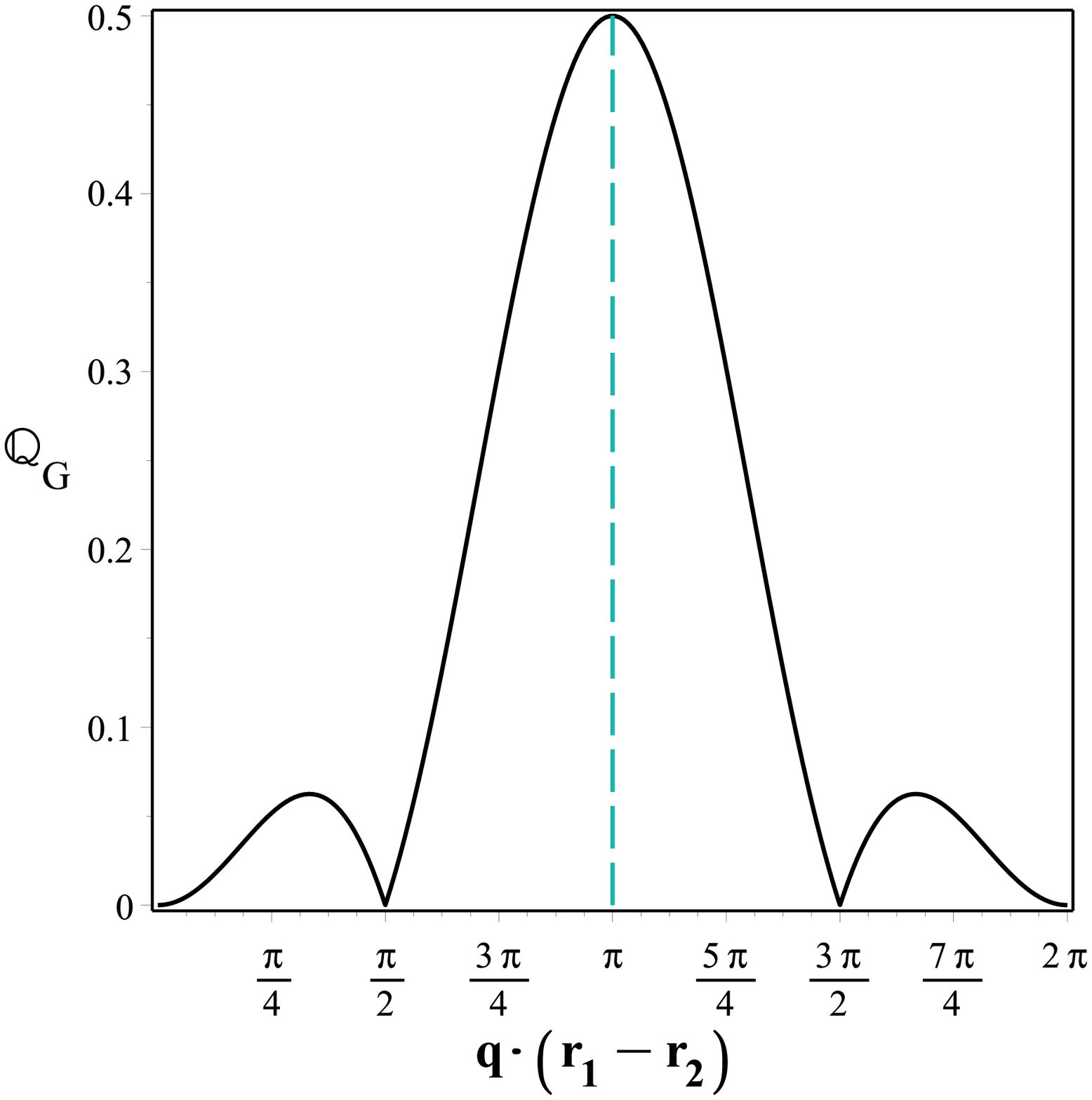}}\qquad
\subfigure[]{\includegraphics[scale=0.33]{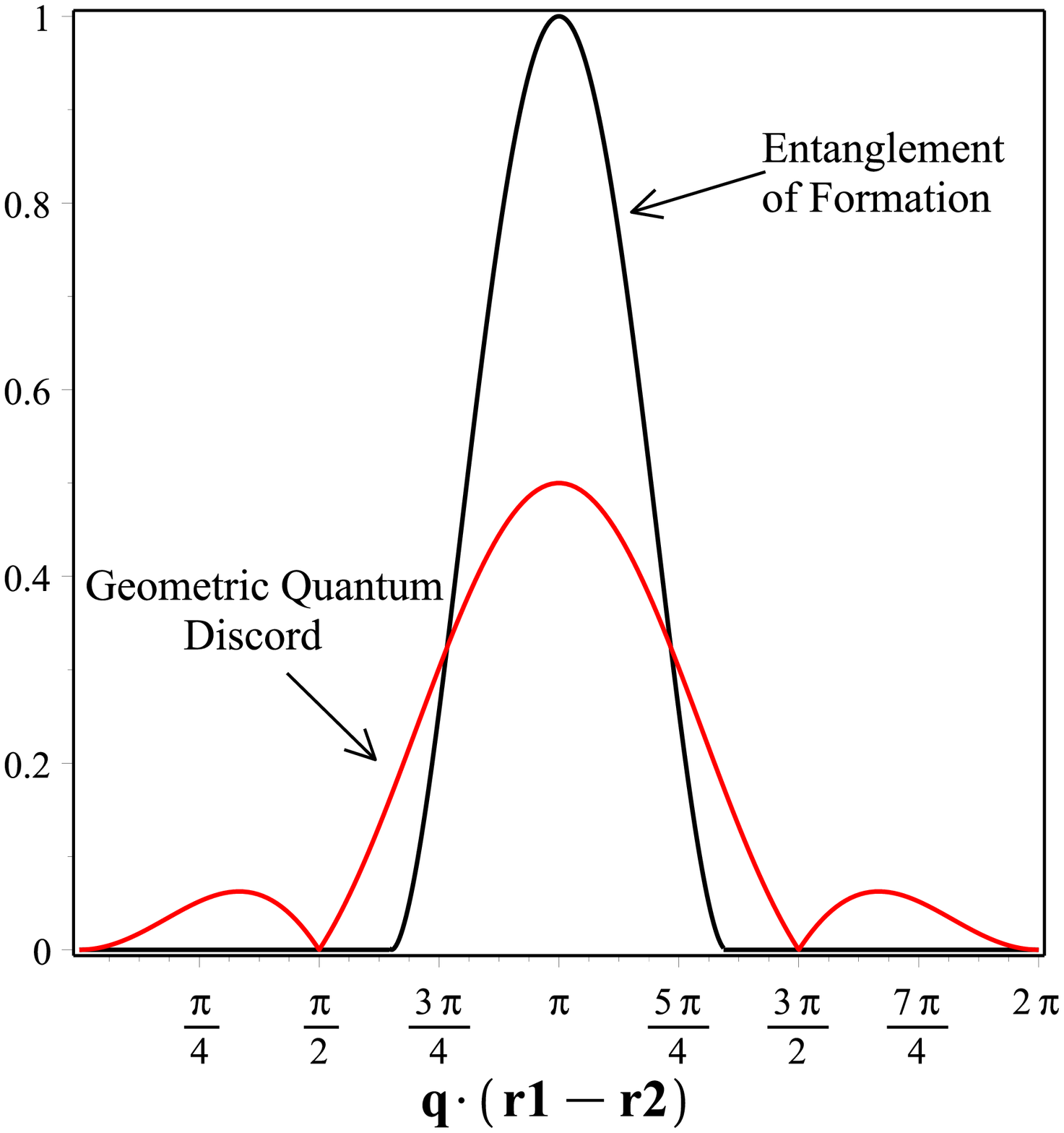}}
\caption{(a) Geometric quantum discord as a function of $\vec{q}\cdot\left(\vec{r_1} - \vec{r_2}\right)$. Dashed (green) vertical line highlights the point where the system is in an antiparallel magnetic alignment with an entangled pure state. (b) We make a comparison between the geometric quantum discord and the entanglement of formation{;} it is possible to identify {the} presence of quantum correlations when the entanglement is absent and even when the system is found in a parallel magnetic alignment with a separable quantum state.}
\label{fig:geometric}
\end{figure}

Fig. \ref{fig:geometric} (a) shows the geometric quantum discord as a function of the scattering vector $\vec{q}$ times the distance between the metallic centers $\vec{r_1} - \vec{r_2}$ of a sample material, {Eq.(\ref{geometric})}. We identify a maximum of quantum correlation at $ r_1 - r_2 = \pi q^{-1} $, this is compatible to the previous results, where at this point the system is found in {an} entangled pure state, see Fig. \ref{fig:entanglement} (a). {Fig. \ref{fig:geometric} (b)} makes a comparison between the geometric quantum discord and the entanglement of formation. As can be seen, it is possible to identify {the} presence of quantum correlations when the entanglement is absent and even when the system is found in a parallel magnetic alignment with a separable quantum state ($0 < \vec{q} \cdot \left( r_1 - r_2 \right) < \pi/2 $ and $3\pi /2 < \vec{q} \cdot \left( r_1 - r_2 \right) < 2\pi $ ), see Fig. \ref{fig:spin}; furthermore, the points of zero discord {coincide} with the points of zero correlation ($\pi/2$ and $3\pi /2$), indicating the absence of magnetic interaction ($J=0$) between the magnetic ions.

Therefore, we provide one way to find the geometric quantum correlations in a Heisenberg spin dimer, by diffractive properties obtained via neutron scattering experiments, without {making} any assumption about their macroscopic quantities.

\section{Conclusions}

In summary, our main result was to provide to the literature analytical expressions for quantum information quantifiers, such as {the} entanglement witness, entanglement of formation, Bell's inequality violation and geometric quantum discord as a function of quantities typically obtained in neutron scattering via scalar structure factor. We provide one path toward {identifying} the presence of quantum correlations {and quantum nonlocality} in a two-qubit system {such} as a Heisenberg spin-$1/2$ dimer, using diffractive properties without {making} any {assumptions} about their macroscopic quantities. We present an alternative way to describe the quantum properties of a sample material via neutron scattering experiments. Our results {open} doors for the detection and manipulation of quantum correlations through neutron scattering experiments in magnetic systems{, such as the} molecular magnets ruled by Heisenberg Hamiltonians; leading to promising applications in quantum information science, since these materials can be promising platforms in quantum information processing. 

\begin{acknowledgments}
The author would like to thank D. O. Soares-Pinto and M. S. Reis for helpful comments, and specially, E. E. M. Lima for the computational help. This work was supported by Brazilian funding agencies CNPq, CAPES and FAPERJ.
\end{acknowledgments}

\end{document}